\documentclass[12pt]{elsarticle}

\usepackage{mathstyle00}

\DeclareMathOperator{\Tr}{Tr}

\makeatletter
\def\ps@pprintTitle{%
  \let\@oddhead\@empty
  \let\@evenhead\@empty
  \let\@oddfoot\@empty
  \let\@evenfoot\@oddfoot
}
\makeatother

\usepackage[margin=1.0in]{geometry}    % Margin: 1-inch all round
\usepackage{caption}
\usepackage{tikz}
\usetikzlibrary{matrix,shapes,arrows,positioning}
\begin{document}
\begin{frontmatter}
		\title{A mesh-free method for interface problems using the deep learning approach}
		
		\author[hku]{Zhongjian Wang}
		\ead{ariswang@connect.hku.hk}
		\author[hku]{Zhiwen Zhang\corref{cor1}}
		\ead{zhangzw@hku.hk}
		
		\address[hku]{Department of Mathematics, The University of Hong Kong, Pokfulam Road, Hong Kong SAR, China.}
		\cortext[cor1]{Corresponding author}
\begin{abstract}
In this paper, we propose a mesh-free method to solve interface problems using the deep learning approach. Two interface problems are considered. The first one is an elliptic PDE with a discontinuous and high-contrast coefficient. While the second one is a linear elasticity equation with discontinuous stress tensor. In both cases, we formulate the PDEs into variational problems, which can be solved via the deep learning approach.
To deal with the inhomogeneous boundary conditions, we use a shallow neuron network to approximate the boundary conditions. Instead of using an adaptive mesh refinement method or specially designed basis functions or numerical schemes to compute the PDE solutions, the proposed method has the advantages that it is easy to implement and mesh-free. Finally, we present numerical results to demonstrate the accuracy and efficiency of the proposed method for interface problems. \\

% 68T99    Computer science  Artificial intelligence  None of the above, but in this section
% 35R05    Partial differential equations with discontinuous coefficients or data
% 35J20    Variational methods for second-order elliptic equations
% 74B05  	Classical linear elasticity
% 65N30  	Finite elements, Rayleigh-Ritz and Galerkin methods, finite methods

\end{abstract}
		
\end{frontmatter}

\section{Introduction}
\noindent
In recent years, deep learning methods have achieved unprecedented successes in various application fields, including computer vision, speech recognition, natural language processing, audio recognition, social network filtering, and bioinformatics, where they have produced results comparable to and in some cases superior to human experts \cite{lecun2015deep,goodfellow2016deep}. Motivated by these exciting progress, there are increased new research interests in the literature  for the application of deep learning methods for scientific computation, including approximating 
multivariate functions and solving differential equations using the deep neural network; see 
\cite{JinchaoXu:2018,QiangDU:2018,weinan2017deep,weinan2018deep,khoo2017solving,ZabarasZhu:2018} and references therein. 

In \cite{JinchaoXu:2018}, the authors investigate the relationship between deep neural networks with rectified linear unit (ReLU) function as the activation function and continuous piecewise linear functions in the finite element method (FEM). A new error bound for the approximation of multivariate functions using deep ReLU networks is presented in \cite{QiangDU:2018}, which shows that the curse of the dimensionality is lessened by establishing a connection between the deep networks and sparse grids. In \cite{weinan2018deep} the authors solve Poisson problems and eigenvalue problems in the context of the Ritz method based on representing the trail functions by deep neural networks. Meanwhile, in \cite{weinan2017deep} the authors propose deep learning-based numerical methods for solving high-dimensional parabolic partial differential equations and backward stochastic differential equations. In \cite{khoo2017solving}, a neural network was proposed to learn the physical quantity of interest as a function of random input coefficients; the accuracy and efficiency of the approach for solving parametric PDE problems was shown. In \cite{ZabarasZhu:2018}, the authors propose a Bayesian approach to develop deep convolutional encoder-decoder networks, which give surrogate models for uncertainty quantification and propagation in problems governed by stochastic PDEs. In \cite{Yalchin2018deep}, the authors design multi-layer neural network architectures for multiscale simulations of flows that takes into account the observed data and physical modeling concepts. In \cite{schwab2017deep}, the authors estimate the expressive power of a class of deep Neural Networks on a class of countably-parametric maps. Those maps arise as response surfaces of parametric PDEs with distributed uncertain inputs. 
 
In this paper, we investigate the deep learning approach to solve interface problems, which have many application in physical and engineering sciences. For example, to model the heterogeneous porous medium in the reservoir simulation, the permeability field is often assumed to be a multiscale function with high-contrast and discontinuous features. Another example is to study the evolution of the shape and location of fibroblast cells under stress \cite{zhang2018minimal}. The model is based on ideas of a continuum mechanical description of stress-induced phase transitions, where the cell is modeled as a transformed inclusion in a linear elastic matrix and the cell surface  evolves according to a special kinetic relation. In this model, the stress tensor has discontinuity across the cell surface due to the transformation in the strain tensor caused by contraction in the cell. 

%Many numerical methods were proposed in the literature to solve PDEs with discontinuous coefficients. 
There has been a lot of effort in developing accurate and efficient finite element methods (FEMs) for interface problems. In \cite{li2003new,gong2008immersed}, Li et.al. developed the immersed-interface finite element method to solve elliptic interface problems with non-homogeneous jump conditions. Their method considered uniform triangular grids and approximated the interface by a straight line segment when it intersects a coarse grid element. By matching the jump condition, they created a special basis function for elements which were cut through by the interface and proved a second order convergence rate in the $L_2$ norm and a first order convergence rate in the $H_1$ semi-norm. However, the constants in their error estimate depend on the contrast of the coefficient. In \cite{GrahamHou:10},  Hou et.al. developed a new multiscale finite element method which was able to accurately capture solutions of elliptic interface problems with high-contrast coefficients by using only coarse quasi-uniform meshes, and without resolving the interfaces. Moreover, they provided optimal error estimate in the sense that the hidden constants in the estimates were independent of the contrast of the PDE coefficients. Much earlier, Babu\v{s}ka \cite{babuvska1970finite} studied the convergence of methods based on a minimization problem equivalent to elliptic PDEs with discontinuous coefficients, in which the boundary and jump condition were incorporated in the cost functions. In \cite{chen1998finite}, Chen and Zou approximated the smooth interface by a polygon and used classical finite element methods to solve both elliptic and parabolic interface equations, where the mesh must align with the interface. 
 
Alternatively, some efficient finite difference methods (FDMs) were proposed to solve interface problems. 
In \cite{peskin1977numerical}, Peskin developed the immersed boundary method (IBM) to study the motion of one or more massless, elastic surfaces immersed in an incompressible, viscous fluid, particularly in bio-fluid dynamics problems where complex geometries and immersed elastic membranes are present. The IBM method employs a uniform Eulerian grid over the entire domain to describe the velocity field of the fluid and a Lagrangian description for the immersed elastic structure. We refer to \cite{peskin2002immersed} for an extensive review of this method and its various applications. Another related work is the immersed interface method (IIM) for elliptic interface problems
developed by LeVeque and Li \cite{leveque1994immersed}. By incorporating the jump condition across the interface to modify the finite difference approximation near the interface, a second order accuracy was maintained. 
An important development of interface capturing methods is the ghost fluid method (GFM) developed by Osher et.al.\cite{fedkiw1999non}, which incorporated the interface jump condition into the finite difference discretization by tracking the interface with a level set function. The GFM has been applied to capture discontinuities in multi-medium compressible multiphase flows.

%Instead of computing on a uniform grid, other approaches use a structured grid or adaptive mesh that is deformed in neighborhood of the discontinuity to efficiently solve the PDEs; see \cite{berger1984adaptive,berger1989local,thompson1985numerical,chen1998finite,ceniceros2001efficient,babuska2012modeling,tang2003adaptive,ainsworth1997posteriori} and references therein. For instance, in \cite{chen1998finite} Chen and Zou approximated the smooth interface by a polygon and used classical finite element methods to solve both elliptic and parabolic interface equations, where the mesh must align with the interface. 

In this paper, we are interested in developing numerical methods to solve interface problems in a mesh-free manner. Our work is inspired by the deep Ritz method proposed in \cite{weinan2018deep}, where the Poisson problems and eigenvalue problems were studied. We intend to investigate the expressive power of the deep neural networks in representing solutions of interface problems. Two typical interface problems are considered. The first one is an elliptic PDE with a discontinuous and high-contrast coefficient, which is a challenging problem and has been intensively studied; see \cite{bernardi2000adaptive,li2003new,GrahamHou:10,efendiev2011multiscale}. The second one is a linear elasticity equation with discontinuous stress tensor \cite{zhang2018minimal}.
%which can be solved using the immersed interface method \cite{LiLi2003elasticity} or matched interface and boundary method \cite{wang2015matched}. 

In both problems, we formulate the PDEs into variational problems, which can be solved using the deep learning approach. Then, we use the stochastic gradient descent (SGD) method to solve the variational problem. 
To impose inhomogeneous boundary conditions, we propose to use a shallow neuron network to approximate the boundary conditions. We find that the proposed method is easy to implement and mesh-free since we do not need to choose an adaptive mesh to discretize the PDEs. Our numerical results show that the proposed method can efficiently solve the interface problems. Moreover, we observe that the convergence time of the SGD method is random, which may be due to the fact that the iteration process of the SGD method can be get stuck into some local minimums. Especially, we find that it takes a longer time to get out of local minimums in a `harder' case of the high-contrast problem; see Section \ref{sec:NumExaEllipticProblem} for more details.

%\textcolor{red}{??? some discussions about the pros and cons of our method.} \textcolor{red}{??? Brief conclusion of the numerical results.} 
The rest of the paper is organized as follows. In Section 2, we shall review the basic ideas of deep neural network and the idea of the deep Ritz method.  In Section 3, we propose the formulation of the deep learning method in solving interface problems. We also discuss the issues regarding the implementation of the proposed method, including how to impose inhomogeneous boundary conditions. In Section 4, we present numerical results to demonstrate the accuracy of our method. Concluding remarks will be made in Section 5.
\section{Some preliminaries}
\noindent
In this section, we briefly discuss the definition and properties of the deep neural network (DNN), including its approximation property and then the formulation of the deep Ritz method \cite{weinan2018deep}.

\subsection{The DNN and its approximation property}\label{subsec:DNN}
\noindent
There are two ingredients in defining a DNN. The first one is a (vector) linear function of the form $T:R^n\rightarrow R^m$, defined as $T(x)=Ax+b$, where $A=(a_{ij})\in R^{m\times n}$, $x\in R^{n}$ and $b\in R^m$. The second one is a nonlinear activation function $\sigma:R\rightarrow R$. A frequently used activation fucntion, known as the rectified linear unit (ReLU), is defined as $\sigma(x)=\max(0,x)$ \cite{lecun2015deep}. In the artificial neural network literature, the Sigmoid function is another frequently used activation function, which is defined as $\sigma(x)=(1+e^{-x})^{-1}$. By applying the activation function in 
an element-wise manner, one can define (vector) activation function $\sigma: R^n\rightarrow R^n$.

Equipped with those definitions, we are able to define a continuous function $F(x)$ by a composition of 
linear transforms and activation functions, i.e., 
\begin{equation}\label{eqn:eg3layernet}
F(x)=T^{k}\circ\sigma\circ T^{k-1}\circ\sigma
\cdot\cdot\cdot\circ T^{1}\circ\sigma\circ T^{0}(x),
\end{equation}
where $T^{i}(x)=A_ix+b_i$ with $A_i$ be undetermined matrices and $b_i$ be undetermined vectors, and $\sigma(\cdot)$ is the element-wisely defined activation function. Dimensions of $A_i$ and $b_i$ are chosen to make \eqref{eqn:eg3layernet} meaningful. Such a DNN is called a $(k+1)$-layer DNN, which has $k$ hidden layers. Denoting all the undetermined coefficients (e.g., $A_i$ and $b_i$) in \eqref{eqn:eg3layernet} as $\theta\in\Theta$, where $\theta$ is a high dimensional vector and $\Theta$ is the space of $\theta$. The DNN representation of a continuous function can be viewed as 
\begin{align}\label{eqn:solution_DNN}
F=F(x;\theta).
\end{align}
Let $\mathbb{F}=\{ F(\cdot,\theta)|\theta\in\Theta\}$ denote the set of all expressible functions by the DNN parametrized by $\theta\in\Theta$. Then $\mathbb{F}$ provides an efficient way to represent unknown continuous functions, comparing with a linear solution space used in classic numerical methods, e.g., a trial space spaced by linear nodal basis functions in the FEM. In the sequel, we shall discuss the approximation property of the DNN, which is relevant to the study of the expressive power of a DNN model \cite{cohen2016expressive,schwab2017deep}. 

Early studies of approximation properties of neural network can be found in \cite{cybenko1989approximation,hornik1989multilayer}, where the authors studied approximation properties for the function classes given by a feed-forward neural network with a single hidden layer.  Later, many authors studied the error estimates for such neural networks in terms of number of  neurons, layers of the network, and activation functions; see \cite{ellacott1994aspects,pinkus1999approximation} for a good review of relevant works.

In recent years, the DNN has shown successful applications in a broad range of problems, including classification for complex systems and construction of response surfaces for high-dimensional models. Significant efforts have been devoted to study the benefits on the expressive power of NNs afforded by NN depth. For example, in \cite{cohen2016expressive}, the authors proved that convolutional DNNs were able to express multivariate functions given in so-called Hierarchic Tensor (HT) formats. In \cite{yarotsky2017error}, the author studied the expressive power of shallow and deep neural networks with piece-wise linear activation functions and established new rigorous upper and lower bounds for the network complexity in approximating Sobolev spaces. 

In \cite{JinchaoXu:2018}, the authors studied the relationship between DNNs with ReLU function as the activation function and continuous piecewise linear functions from the linear FEM. 
They proved the following statement. 

\begin{proposition}\label{fem-dnn}
Given a locally convex finite element grid ${\cal T}_h$, any linear finite element function with $N$ degrees of freedom, can be written as a ReLU-DNN with at most $k = \lceil\log_2 k_h\rceil+1$ hidden layers and at most $\mathcal{O}(k_hN)$ number of the neurons, where $k_h$ denotes the maximum number of neighboring elements of one node.
\end{proposition}

The Prop.\ref{fem-dnn} provides upper bounds in setting the number of hidden layers and number of neurons 
within each layer, when one uses the DNN to approximate the solution space spanned by the FEM basis. 
In our numerical results, we find that choosing a relatively small number of hidden layers and neurons 
are good enough to obtain accurate numerical results.

\subsection{Formulation of the deep Ritz method}\label{sec:deepRitz}
\noindent
The deep Ritz method is a deep learning based numerical method for solving variational problems \cite{weinan2018deep}. Therefore, it naturally can be used to solve PDEs. For example, we consider a Poisson equation defined on a compact domain $D\subsetneq R^d$,
\begin{equation}\label{eqn:poisson}
\begin{cases}
-\Delta u(x)=f(x),& \quad x\in D,\\
u(x)=0, &\quad x\in\partial D.
\end{cases}
\end{equation}
Given the Poisson equation \eqref{eqn:poisson}, we can derive the corresponding variational problem as 
\begin{equation}\label{eqn:poisson_variationalform}
J(v)=\frac{1}{2}\int_D \nabla v(x)\cdot \nabla v(x) dx -\int_D v(x)f(x) dx, \quad v\in \mathbb{H}^1_0(D).
\end{equation}
Then, the solution of \eqref{eqn:poisson} can be obtained by,
\begin{equation}
u=\argmin_{v\in \mathbb{H}^1_0(D)}J(v).
\end{equation}
From the perspective of scientific computing, the Poisson equation \eqref{eqn:poisson} can be solved using numerical methods, such as FDMs and FEMs. From the perspective of machine learning however, the numerical solution of $u(x)$ is interpreted as a function with $x\in R^d$ as its input and $R^1$ as its output, where $d$ denotes the dimension the physical domain $D$. Thus, it can be approximated by $F(x)$ in \eqref{eqn:eg3layernet}. 

Let $\tilde{u}$ denote the DNN representation of the solution of the Poisson equation. Substituting $\tilde{u}$ into the variational problem \eqref{eqn:poisson_variationalform}, we get the optimization problem 
\begin{equation}\label{eqn:lag_representation}
\tilde{u}=\argmin_{F\in \mathbb{F}_0}J(F),
\end{equation}
where $\mathbb{F}_0$ is a subspace of $\mathbb{F}$ that satisfies the boundary condition on $\partial D$ and it may have some limitations on imposing boundary conditions. The justification of this assumption will be discussed later.  

%Notice that the geometry of the space $\mathbb{F}$ is complicated. 
After parameterizing the expressible function space by $\theta\in\Theta$, we equivalently define the 
variational problem \eqref{eqn:poisson_variationalform} as  
\begin{equation}\label{eqn:lag_representation_para}
 \min_{\theta\in\Theta}J(\theta) = 
\frac{1}{2}\int_D  |\nabla F(x,\theta)|^2 dx -\int_D F(x,\theta)f(x) dx.
\end{equation}
The variational problems \eqref{eqn:lag_representation_para} is not convex in general even when the original variational problem \eqref{eqn:poisson_variationalform} is. In other word, the variational problem \eqref{eqn:poisson_variationalform} is convex with respect to the solution $u(x)$, however, the variational problem \eqref{eqn:lag_representation_para} is non-convex with respect to the parameters in the DNN. Obviously, the issue of local minima and saddle points is nontrivial, which brings essential challenges to many existing optimization methods. 

Since the parameter space $\Theta$ is typically very large, one usually uses the stochastic gradient descent (SGD) method \cite{bottou2010large} to solve \eqref{eqn:lag_representation_para}. There are plenty of optimization methods to search among the large parameter space. To accelerate the training of the neural network, we use the Adam optimizer version of the SGD \cite{kingma2014adam}.

To impose boundary conditions is an important issue in the DNN representation. 
In the homogeneous Dirichlet problem \eqref{eqn:poisson}, a relaxation approach was proposed 
to address this issue. Specifically, one adds a soft constraint (a boundary integral term) to the 
functional $J(\cdot)$ defined in \eqref{eqn:lag_representation_para} and obtains 
\begin{equation}\label{eqn:lag_representation_bdd}
\tilde{u}_\epsilon=\argmin_{F\in \mathbb{F}}\Big(J(F)+\frac{1}{\epsilon}\int_{\partial D}F(x,\theta)^2 dx\Big).
\end{equation} 
Notice that the soft constraint term $\frac{1}{\epsilon}\int_{\partial D}F(x,\theta)^2 dx$ will approach zero when we decrease the parameter $\epsilon$ in the calculation. Therefore, the homogeneous boundary condition is satisfied in a certain weak scene. 
 
\section{Inhomogeneous boundary condition}\label{sec:InhomogeneousBC}
\noindent
As an extension to the deep Ritz method, we consider to solve the inhomogeneous Dirichlet problem as follows
\begin{equation}\label{InhomogeneousProblem}
\begin{cases}
\mathcal{L}u(x)=f(x),\quad x\in D,\\
u(x)=g(x),\quad x\in \partial D,
\end{cases}
\end{equation}
where $\mathcal{L}$ is a linear PDE operator, $f(x)$ is a source function, and $g(x)$ is a boundary condition. 
Let $J(v;f)$ denote the Lagrangian form associated with the homogeneous Dirichlet problem of \eqref{InhomogeneousProblem}, i.e., $g(x)=0$; see \eqref{eqn:poisson_variationalform} for instance.

To deal with the inhomogeneous boundary condition in \eqref{InhomogeneousProblem}, we first choose a shallow neuron network to approximate the boundary condition $g(x)$. Let $\tilde{g}(x)$ denote the approximation 
of $g(x)$ using the neuron network, which is defined on whole domain $D$. However, only boundary values of $\tilde{g}$ are used, so it can be obtained by solving the following optimization problem
\begin{equation}\label{optimzationinhomoBC}
\tilde{g}(x) = \argmin_{G\in \mathbb{G}}\Big(\int_{\partial D}\big(G-g(x)\big)^2dx\Big),
\end{equation}
where $\mathbb{G}$ denotes the set of all expressible functions by a shallow neuron network. 
The optimization problem \eqref{optimzationinhomoBC} can be approximated by,
\begin{equation}\label{eqn:tilde_g_num}
\frac{vol(\partial D)}{N_1}\sum_{i=1}^{N_1}\big(G(y_i)-g(y_i)\big)^2,
\end{equation}
where $y_i\overset{i.i.d.}{\sim} Unif(\partial D)$ and $N_1$ is the number of sample points. In real application, uniform sampler of $\partial D$ is not necessary. One can change the integrand of \eqref{optimzationinhomoBC} by multiplying the Radon-Nikodym derivative of the sampler's distribution. 
Once we obtain a sampler whose distribution is absolutely continuous w.r.t Lebesgue measure of $\partial D$, we can still minimizing \eqref{eqn:tilde_g_num} to obtain $\tilde{g}(x)$. 

In our proposed approach, reasons of choosing a shallow network to approximate $g(x)$ are twofold. First,  $\tilde{g}(x)$ plays as the role of an initial guess to the inhomogeneous boundary condition.  As explained above, only the values of $g(x)$ on $\partial D$ will be used, so limited parameters of $\tilde{g}(x)$ will be good enough. This helps shorten the training of $\tilde{g}$.  Second, due to the simple structure of $\tilde{g}$, the term $\mathcal{L}\tilde{g}\cdot v$ in $J(v;f-\mathcal{L}\tilde{g})$ will not oscillate in $D$ (especially in the weak form), which leads to a faster convergence in solving optimization problems. 
 
\begin{figure}
	\begin{center}
		\begin{tikzpicture}[->,>=stealth',shorten >=1pt,auto,node distance=3cm,
		thick,main node/.style={circle,fill=blue!20,draw,
			font=\sffamily,align=center}]		
		\node[main node] (x) {Input\\ $x$};
		\node[main node] (l1) [right of=x] {Layer 1,\\$w=10$};
		\node[main node] (l2) [right of=l1] {Layer 2,\\$w=10$};
		\node[main node] (l3) [right of=l2] {Layer 3,\\$w=10$};
		\node[main node] (g) [right of=l3] {Output\\ $\tilde{g}(x)$};
		
		\path[every node/.style={font=\sffamily\scriptsize,
			fill=white,inner sep=1pt}]
		% Right-hand-side arrows rendered from top to bottom to
		% achieve proper rendering of labels over arrows.
		(x) edge [bend left=60] node[above=1mm] {Linear+Activation} (l1)
		edge [bend right=40] node[below=1mm] {Linear+Activation} (l2)
		(l1) edge [bend left=60] node[above=1mm] {Linear+Activation} (l2)
		(l2) edge [bend left=60] node[above=1mm] {Linear+Activation} (l3)
		(l3) edge [] node[above=1mm] {Linear} (g);
		\end{tikzpicture}
	\end{center}
	\caption{Network Layout for $\tilde{g}$. }
	\label{twoDNNs_g}
%\end{figure}
%\begin{figure}
	\begin{center}
		\begin{tikzpicture}[->,>=stealth',shorten >=1pt,auto,node distance=3cm, thick,main node/.style={circle,fill=blue!20,draw, font=\sffamily\small,align=center}]
		\node[main node] (x) {Input\\ $x$};
		\node[main node] (l1) [right of=x] {Layer 1,\\$w=15$};
		\node[main node] (l2) [right of=l1] {Layer 2,\\$w=15$};
		\node[main node] (l3) [right of=l2] {Layer 3,\\$w=15$};
		\node[main node] (l4) [right of=l3] {Layer 4,\\$w=15$};
		\node[main node] (u) [right of=l4] {Output\\ $u'(x)$};
		
		\path[every node/.style={font=\sffamily\scriptsize,fill=white,inner sep=1pt}]
		% Right-hand-side arrows rendered from top to bottom to
		% achieve proper rendering of labels over arrows.
		(x) edge [bend left=60] node[above=1mm] {Linear+Activation} (l1)
		edge [bend right=40] node[below=1mm] {Linear+Activation} (l2)
		(l1) edge [bend left=60] node[above=1mm] {Linear+Activation} (l2)
		edge [bend right=40] node[below=1mm] {Linear+Activation} (l3)
		(l2) edge [bend left=60] node[above=1mm]  {Linear+Activation} (l3)
		edge [bend right=40] node[below=1mm] {Linear+Activation} (l4)
		(l3) edge [bend left=60] node[above=1mm] {Linear+Activation} (l4)
		(l4) edge [] node[above=1mm] {Linear} (u);
		\end{tikzpicture}			
	\end{center}
	\caption{Network Layout for $u'$. }
	\label{twoDNNs_u}
\end{figure}

Fig.\ref{twoDNNs_g} and Fig.\ref{twoDNNs_u} show the network layouts for approximating $\tilde{g}$ and $u'$, respectively, where $w$ denotes the width of each hidden layer. For example, Layer 2 in Fig.\ref{twoDNNs_g} is in $\mathbb{R}^{10}$. To be more precise, denote Layer 1 to be $l_1$, Layer 2 to be $l_2$, then,
\begin{equation}
l_2=\sigma(A[l_1;x]+b),
\end{equation}
where $A$ is a $10\times 12$ matrix and $b$ is a $\mathbb{R}^{10}$ vector to be determined.

%Besides the structure introduced in Fig.\ref{twoDNNs_g} and Fig.\ref{twoDNNs_u}, we randomly generate $2048$ points in interior domain $D$ and $256$ points on boundary $\partial D$ every $10$ SGD training steps to calculate approximated integral in \eqref{eqn:lag_representation_bdd}, \eqref{eqn:tilde_g_num}. Learning rate $\eta$ in \eqref{eqn:sgd_update} is chosen as $5\times 10^{-4}$. Once we have a uniform sampler of $D$, the network automatically deals with the interface without knowing locations of the interfaces a-priori. 

Since the neuron network that is used to represent $\tilde{g}$ is shallow, i.e., $\tilde{g}$ is represented by 
a composition of smooth functions, $\mathcal{L}\tilde{g}$ is expressible.
Then, we solve an auxiliary PDE as follows,
\begin{equation}\label{meth:nonzero-diri}
\begin{cases}
\mathcal{L}u^{\prime}(x)=f(x)-\mathcal{L}\tilde{g}(x),\quad x\in D,\\
u^{\prime}(x)=0,\quad x\in \partial D.
\end{cases}
\end{equation}
Now the problem \eqref{meth:nonzero-diri} becomes a homogeneous Dirichlet problem, which can be solved using 
the deep Ritz approach; see Section \ref{sec:deepRitz}. Finally, the solution of the inhomogeneous Dirichlet problem \eqref{InhomogeneousProblem} can be represented as $u(x)=u^{\prime}(x)+\tilde{g}(x)$.   
  
\section{Derivation of the methodology}\label{sec:Derivation}
\subsection{Elliptic PDEs with discontinuous and high-contrast coefficients}\label{sec:DerivationElliptic}
\noindent
We first consider  elliptic PDEs with  discontinuous  coefficients defined as follows, 
\begin{align}
\mathcal{L}(x) u(x) &\equiv -\nabla\cdot(a(x)\nabla u(x))  = f(x), \quad x\in D, \label{EllipticIP_eq}, \\
u(x)&= 0,  \quad x\in \partial D, \label{EllipticIP_bc}
\end{align}
where $D\subseteq R^{d}$ is a bounded spatial domain and the boundary of $D$ is a convex polygon. 
For notation simplification, we first study a homogeneous Dirichlet problem. 
The elliptic PDEs with inhomogeneous boundary conditions can be solved by using the approach studied in Section
\ref{sec:InhomogeneousBC}.

The coefficient $a(x)$ is assumed to be a scalar and has jumps across a number of smooth interior interfaces. 
Denoting the inclusions by $D_1$,...,$D_m$ and setting $D_0=D \setminus \bigcup_{i=1}^{m} D_i$, we assume that the coefficient $a(x)$ is piecewise constant with respect to the decomposition $\{D_i,i=0,...,m\}$.  Setting $a_{min}=\min{a(x)|_{D_i}: i=0,...,m}$ and dividing \eqref{EllipticIP_eq} by $a_{min}$, we rescale the problem. Specifically, let $\alpha(x)=\frac{a(x)}{a_{min}}$ denote the re-scaled coefficient, which is piecewise constant with respect to the partition $\{D_i,i=0,...,m\}$ and $\alpha(x)\geq 1$ for all 
$x\in D$. Letting $\alpha_i$ denote the restriction of $\alpha(x)$ to $D_i$, we are interested in studying two types of high-contrast cases, 
\begin{align}
&\text{Case 1}: \quad \min_{i=1,...,m} \alpha_i \gg 1, \quad \alpha_0=1,
 \label{high_contrast_case1}  \\
&\text{Case 2}: \quad \alpha_0 \gg 1, \quad  
\max_{i=1,...,m} \alpha_i \leq K,  \label{high_contrast_case2}
\end{align}
for some positive constant $K$. In Case 1, the inclusions are high permeability compared to
the background, while the Case 2 contains the converse configuration. 
%\textcolor{red}{describe the background of the coefficient and its difficulty??? Thats just a guess... ??} 

Now, we are in the position to derive the formulation of deep learning approach to solve the elliptic PDEs \eqref{EllipticIP_eq}\eqref{EllipticIP_bc} with high-contrast coefficients \eqref{high_contrast_case1} 
\eqref{high_contrast_case2}. We define the corresponding variational problem as 
\begin{equation}\label{eqn:multiscale_variationalform}
J(v)=\frac{1}{2}\int_D   a(x) |\nabla v(x)|^2 dx -\int_D v(x)f(x) dx, \quad v\in \mathbb{H}^1_0(D).
\end{equation} 
Then, the solution of \eqref{EllipticIP_eq}\eqref{EllipticIP_bc} can be obtained by 
$u(x)=\argmin_{v\in \mathbb{H}^1_0(D)}J(v)$, where $J(\cdot)$ is defined in \eqref{eqn:multiscale_variationalform}. Again, we denote the set of all expressible function by  $\mathbb{F}=\{ F(\cdot,\theta)|\theta\in\Theta\}$ and set $\mathbb{F}_0=\{F\in\mathbb{F}\big|F|_{\partial D}=0\}$. Moreover, let $\Theta_0$ denote the parameter set satisfies the homogeneous boundary condition, i.e., $F(\cdot,\theta)|_{\partial D}=0$, $\theta\in\Theta_0$. The approximation property of the DNN implies that $\mathbb{F}_0\subsetneq \mathbb{C}_0^\infty(D)\subsetneq \mathbb{H}_0^1(D)$. Therefore, we represent the solution $u(x)$ to Eq.\eqref{EllipticIP_eq} using the DNN method. 

Let $\tilde{u}=F(x;\theta)$ denote the DNN representation; see Eq.\eqref{eqn:eg3layernet}. Then, $\tilde{u}$ satisfies the following variational problem
\begin{equation}\label{multiscale_goal_deepritz}
\tilde{u}=\argmin_{F=F(\cdot,\theta)|\theta\in \Theta_0}\frac{1}{2}\int_D a(x) |\nabla F(x,\theta)|^2 dx
-\int_D F(x,\theta)f(x) dx.
\end{equation}
Since the degree of freedom in the variational problem \eqref{multiscale_goal_deepritz} is quite large, 
we apply the  SGD method on the parameter space $\Theta_0$ to solve it. As such, we approximate gradient of one parameter $\theta_k$ by,
\begin{align}
\frac{\partial J(F(\cdot,\theta))}{\partial \theta_k}&=\frac{1}{2}\int_D \frac{\partial(a(x)|\nabla F(x,\theta)|^2 )}{\partial\theta_k} dx -\int_D \frac{\partial(Ff)}{\partial\theta_k} dx \nonumber\\
&\approx\frac{vol(D)}{N}\sum_{i=i}^{N}\Big(\frac{1}{2} \frac{\partial(a(x_i)|\nabla F(x_i,\theta)|^2 )}{\partial\theta_k} - \frac{\partial(F(x_i,\theta)f(x_i))}{\partial\theta_k}\Big)\label{num_lag},
\end{align} 
where $x_i\overset{i.i.d.}{\sim} Unif(D)$ are randomly sampled from the physical domain $D$, $vol(D)$ is the volume of the domain, and $N$ is called batch number in the context of deep learning (meaning the number of training examples utilized in one iteration). Notice that $\theta$ is a high-dimensional vector and $\theta_k$ is any component of $\theta$. After we get the approximation of the gradient with respect to $\theta_k$, we can update each component of $\theta$ as 
\begin{equation}\label{eqn:sgd_update}
\theta_k^{n+1} =  \theta_k^{n} - \eta \frac{\partial J(F(\cdot,\theta))}{\partial \theta_k}|_{\theta_k=\theta_k^{n}},
\end{equation} 
where $ \eta$ is the learning rate. To accelerate the training of the neural network, we use the Adam optimizer version of the SGD method \cite{kingma2014adam}.
\begin{remark} 
	From the derivation of the DNN formulation, one can see that the proposed method automatically deals with the interface condition (or discontinuous coefficients) without knowing locations of the interfaces \textit{a-priori}.
\end{remark}

\subsection{Linear elasticity with discontinuous stress tensors} 
\noindent
In this subsection, we consider the DNN approach to solve linear elasticity interface problems.
One application of the linear elasticity problem is to model the shape and location of fibroblast cells under stress \cite{zhang2018minimal}. The model is based on the idea of a continuum mechanical description of stress-induced phase transitions. To demonstrate the main idea, we consider a two-dimensional linear elasticity problem.

Suppose the matrix (meaning the material or tissue in cells) plus the cell together occupy a bounded domain $D \subseteq R^d$, $d=2$ and $D$ is composed of linear elastic homogeneous isotropic material. We assume the cell has small deformations, so that the linearized theory of elasticity is used. Let $\textbf{u}=(u_1,u_2)^{T}$ denote the displacement field. Then, the strain tensor is
\begin{align}
\textbf{E}=\frac{1}{2}(\nabla\textbf{u}+\nabla\textbf{u}^{T}), \text{with} 
\quad 
E_{ij}=\frac{1}{2}\big(\frac{\partial u_i}{\partial x_j}+\frac{\partial u_j}{\partial x_i}\big). 
\end{align}
In the matrix except the cell, the stress tensor is related to the strain tensor 
(gradient of the displacement) by $\textbf{S}=\mathbb{C}\textbf{E}=\mathbb{C}\nabla\textbf{u}$, where the elasticity tensor $\mathbb{C}$ is a linear transformation on the tensors. In the isotropic case, we have
\begin{align}\label{ElasticityTensor1}
\mathbb{C}\textbf{A}=\lambda \Tr(\textbf{A})\textbf{1}+\mu(\textbf{A}+\textbf{A}^{T})
\end{align}
for any two dimensional matrix $\textbf{A}$. In Eq.\eqref{ElasticityTensor1}, $\lambda$ and $\mu$ are lam\'{e} constants, $\Tr(\cdot)$ is the trace operator, and $\textbf{1}$ is the identity matrix. In components, the action of the elasticity tensor $\mathbb{C}$ reads
\begin{align}\label{ElasticityTensor2}
C_{ijkl}A_{kl}=\lambda A_{kk}\delta_{ij}+\mu(A_{ij}+A_{ji}), 
\end{align}
where the Einstein summation convention is used.

The cell is modeled by a compact region $\Omega$ with smooth boundary; see Fig.\ref{fig:eg2chi}. Let $\textbf{E}_0$ denote a transformation strain, which is a constant symmetric matrix. We assume the stress tensor has a jump across the cell, i.e., 
\begin{align}\label{strain_JumpCondition}
\textbf{S}=
\left\{
\begin{array}{ll}
\mathbb{C}\textbf{E},    & \quad  \text{in}~ D \setminus \Omega,\\     
\mathbb{C}(\textbf{E}-\textbf{E}_{0}),& \quad \text{in}~  \Omega.   
\end{array}
\right.
\end{align}
In our cell model, we set the transformation strain to be a contraction, which is represented by 
an isotropic compression $\textbf{E}_{0}=-\alpha\textbf{1}$ with $\alpha>0$. 
We suppose the cell model is in a quasi-static state. Therefore, the displacement field $\textbf{u}$ satisfies the following linear elasticity PDE with a discontinuous stress tensor in a weak sense, 
\begin{equation}\label{eqn:compact_elastic}
-\nabla\cdot\big(\mathbb{C}\nabla \textbf{u}-\chi_{\Omega}S_0\big)=0, \quad  x\in D, 
\end{equation}
where $\chi_{\Omega}$ is the characteristic function of the cell domain $\Omega$ and 
$S_0=\mathbb{C}\textbf{E}_{0}$ is a constant symmetric matrix, which measures the effect on the cell boundary
due to the contraction.  We impose Dirichlet boundary conditions on $\partial D$. On the cell boundary $\partial\Omega$, the solution $\textbf{u}$ satisfies the following jump conditions
\begin{equation}\label{eqn:jump_conditions}
[\textbf{u}] = 0, \quad [ \textbf{S}]n = 0,  
\end{equation}
where $n$ is the outward unit normal vector on $\partial\Omega$ and $[\ ]$ denotes the jump across the interface. 

%\sout{Note that in the isotropic case, one has $C_{ijkl}\partial_ku_l=\lambda \partial_ku_k\delta_{ij}+\mu(\partial_ku_l+\partial_lu_k)$ and $C_{ijkl}\textbf{E}_{0,kl}=-\alpha\delta_{ij}(\lambda\delta_{kk}+2\mu)$, where the Einstein summation convention is used.} 

Then, the linear elasticity interface problem \eqref{eqn:compact_elastic}-\eqref{eqn:jump_conditions} can be computed by numerical methods, such as the immersed interface method \cite{LiLi2003elasticity} or matched interface and boundary method \cite{wang2015matched}. However, the implementation of the numerical scheme is not simple due to the jump conditions on the interface, especially when the interface has a complicated geometry.

In the sequel, we shall develop the formulation of solving the linear elasticity interface problem \eqref{eqn:compact_elastic}\eqref{eqn:jump_conditions} using the DNN method. 
In the isotropic case, let $\textbf{e}(\textbf{v})\equiv(e_{ij}(\textbf{v}))$, where 
$e_{ij}(\textbf{v})=\frac{1}{2}(\partial_jv_i+\partial_iv_j)$ and $\textbf{v}=(v_1,v_2)^{T}$ is a 
vector valued function. Then, \eqref{eqn:compact_elastic} is equivalent to,
\begin{align}\label{eqn:compact_elastic2}
-\nabla\cdot(\lambda\Tr(\textbf{e}(\textbf{u}))I_2+2\mu \textbf{e}(\textbf{u})-\chi_{\Omega}S_0)=0.
\end{align}  
Then, the variational problem associated with \eqref{eqn:compact_elastic2} is given by,
\begin{align}\label{eqn:variational_elastic}
J(\textbf{v})=\int_D\Big(\frac{\lambda}{2}\Tr(\textbf{e}(\textbf{v}))^2+\mu \textbf{e}(\textbf{v}):\textbf{e}(\textbf{v})+2\chi_{\Omega}(\lambda+\mu)\Tr(\textbf{e}(\textbf{v}))\Big)dx,
\end{align}
where $:$ denotes the inner product between matrices, i.e., $A:B = \Tr(A^{T}B) = \sum_{i,j} a_{ij}b_{ij}$. Finally, the solution of \eqref{eqn:compact_elastic2} can be obtained by 
$\textbf{u}(x)=\argmin_{\textbf{v}\in (\mathbb{H}^1_0(D))^{2}}J(\textbf{v})$, where $J(\cdot)$ is defined in \eqref{eqn:variational_elastic}. The remaining implementation of the DNN method for \eqref{eqn:variational_elastic} is exactly the same as we discussed in Section \ref{sec:DerivationElliptic}, so we skip the details here.

\section{Numerical Example}\label{sec:NumericalExamle}
\noindent 
In this section, we shall carry out numerical experiments to demonstrate the performance of 
the DNN method in solving interface problems. In addition, we are interested in understanding 
the SGD method in solving the non-convex optimization problem. The TensorFlow  \cite{abadi2016tensorflow} provides an efficient tool to calculate the partial derivatives in \eqref{num_lag}, which will be used in our implementation.
 
\subsection{2D high-contrast elliptic problems}\label{sec:NumExaEllipticProblem}
\noindent
We consider 2D elliptic PDEs with high-contrast coefficients defined as follows, 
\begin{align}
-\nabla\cdot(a(x)\nabla u(x))  &= f(x),  \quad x\in D, \label{NumEllipticIP_eq}  \\
u(x)&= g(x), \quad x\in \partial D, \label{NumEllipticIP_bc}
\end{align}
where $x=(x_1,x_2)$, the domain is $D=[-1,1]\times[-1,1]$, and the coefficient $a(x)$
is a piecewise constant defined by
\begin{equation}\label{NumEllipCoef}
\alpha=\begin{cases}
\alpha_1,\quad r<r_0,\\
\alpha_0,\quad r\geq r_0,
\end{cases}
\end{equation}
where $r=(x^2+y^2)^{1/2}$ and $r_0=\pi/6.28$. Moreover, the source term $f(x)=-9r$ and the boundary condition $g(x)=\frac{r^3}{\alpha_0}+(\frac{1}{\alpha_1}-\frac{1}{\alpha_0})r_0^3$. We choose the source term and 
boundary condition in such a way that the exact solution (in the polar coordinate) is 
\begin{equation}
u(r,\theta)=\begin{cases}
\frac{r^3}{\alpha_1},\quad r<r_0,\\
\frac{r^3}{\alpha_0}+(\frac{1}{\alpha_1}-\frac{1}{\alpha_0})r_0^3,\quad r\geq r_0.
\end{cases}
\end{equation}

\begin{figure}[tbph]
	\centering
	\includegraphics[width=0.6\linewidth]{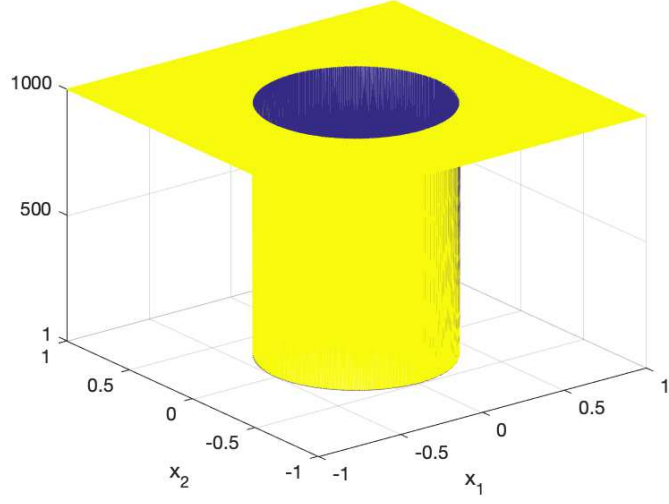}
	\caption{Profile of the high-contrast coefficient $\alpha$ on $D$.}
	\label{fig:eg1alpha3d}
\end{figure}

In our first experiment, we choose $\alpha_0=10^3$ and $\alpha_1=1$ in \eqref{NumEllipCoef}; see Fig.\ref{fig:eg1alpha3d} for the profile of the coefficient. Notice that problem 
\eqref{NumEllipticIP_eq}\eqref{NumEllipticIP_bc} is an inhomogeneous Dirichlet problem. We use the immersed-interface FEM with fine mesh $h=\frac{1}{128}$ to compute the reference solution and the DNN method to compute the numerical solution. The implementation of the DNN method has been intensively discussion in Section \ref{sec:InhomogeneousBC} and Section \ref{sec:DerivationElliptic}. The network that we used is illustrated in Fig.\ref{twoDNNs_u} and Fig.\ref{twoDNNs_g}, which has 4 intermediate layers with width 15 to approximate $u'$ and has 3 intermediate layers with width 10 to approximate $\tilde{g}$. The network is not specially designed for the target problem. Expressibility of DNN discussed in Sec.\ref{subsec:DNN} assures adequate approximation to the solution by adjusting the width of each intermediate layer.  In the learning process, i.e., the running of the SGD method, we choose the batch number (number of samples per gradient update) to be $4352$ (that contains $4096$ points in the interior domain of $D$ and $256$ points on the boundary $\partial D$, which is used to evaluate second term in \eqref{eqn:lag_representation_bdd}) and generate a new batch every $10$ steps of updating. And the learning rate $\eta$ is $5\times 10^{-4}$. Once we have a uniform sampler, the network automatically deals with the interface without knowing locations of the interface a-priori.

\begin{figure}[tbph]
	\centering
	\begin{subfigure}[b]{0.31\textwidth}
		\includegraphics[width=\textwidth]{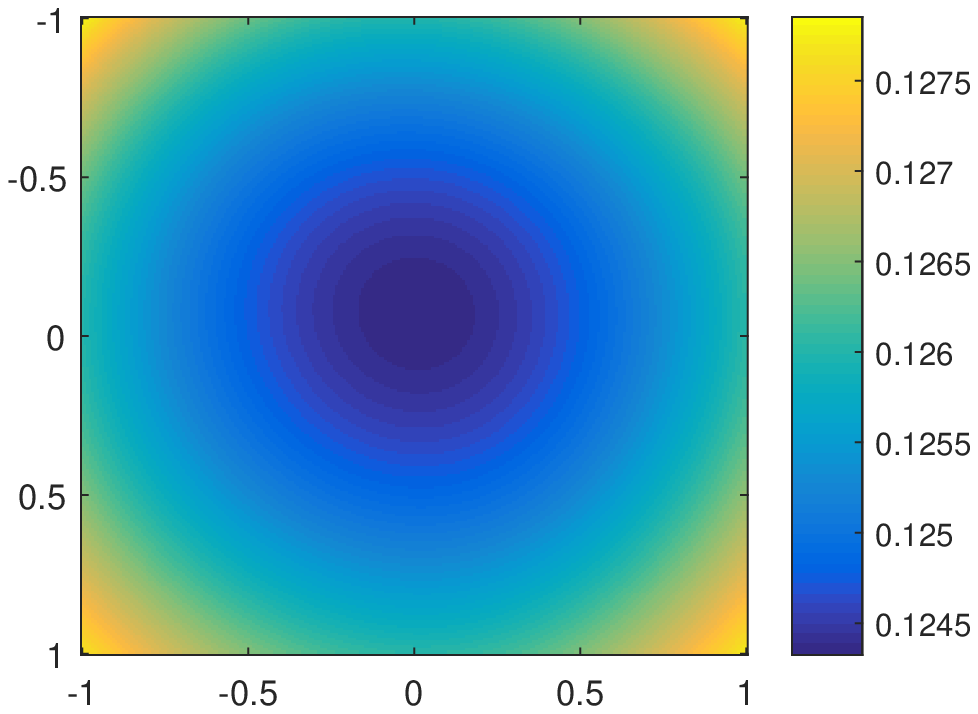}
		\caption{}
		\label{fig:eg1-a}
	\end{subfigure}
	~ %add desired spacing between images, e. g. ~, \quad, \qquad, \hfill etc. 
	\begin{subfigure}[b]{0.31\textwidth}
		\includegraphics[width=\textwidth]{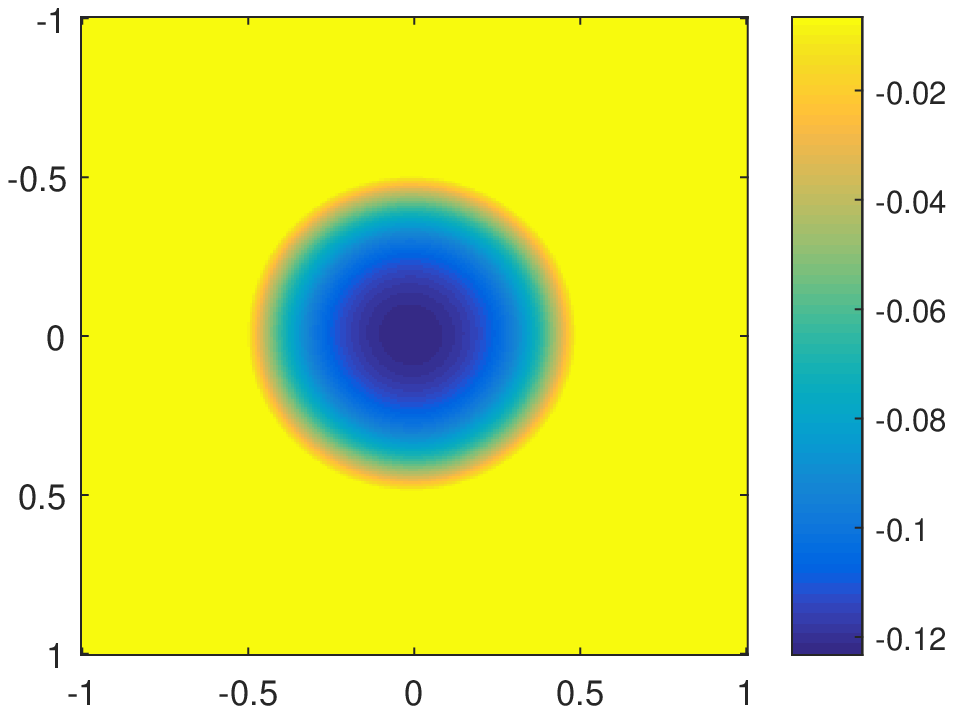}
		\caption{}
		\label{fig:eg1-b}
	\end{subfigure}
	~ 
	\begin{subfigure}[b]{0.31\textwidth}
		\includegraphics[width=\textwidth]{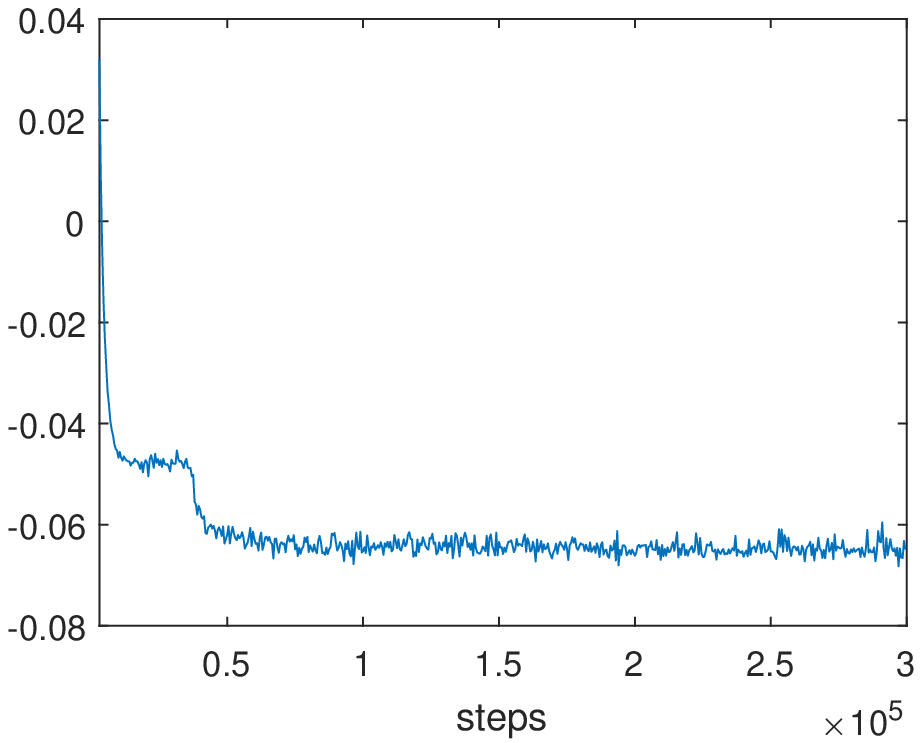}
		\caption{}
		\label{fig:eg1-c}
	\end{subfigure} \\ 
	\centering
    \begin{subfigure}[b]{0.31\textwidth}
	\includegraphics[width=\textwidth]{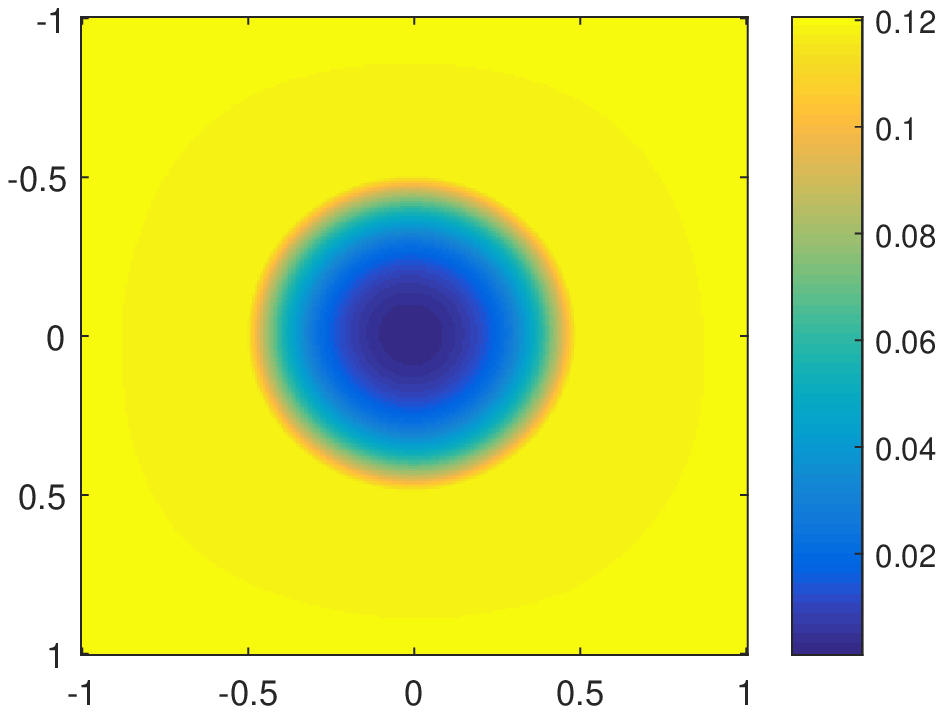}
	\caption{}
	\label{fig:eg1-d}
    \end{subfigure}
    ~ %add desired spacing between images, e. g. ~, \quad, \qquad, \hfill etc. 
    \begin{subfigure}[b]{0.31\textwidth}
	\includegraphics[width=\textwidth]{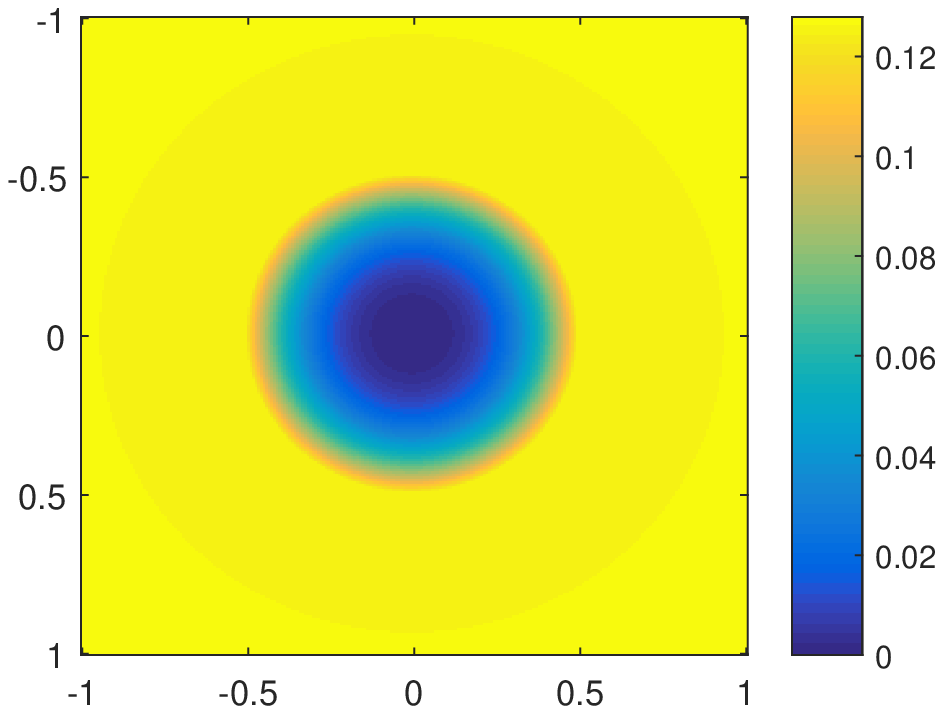}
	\caption{}
	\label{fig:eg1-e}
    \end{subfigure}
    ~ 
    \begin{subfigure}[b]{0.31\textwidth}
	\includegraphics[width=\textwidth]{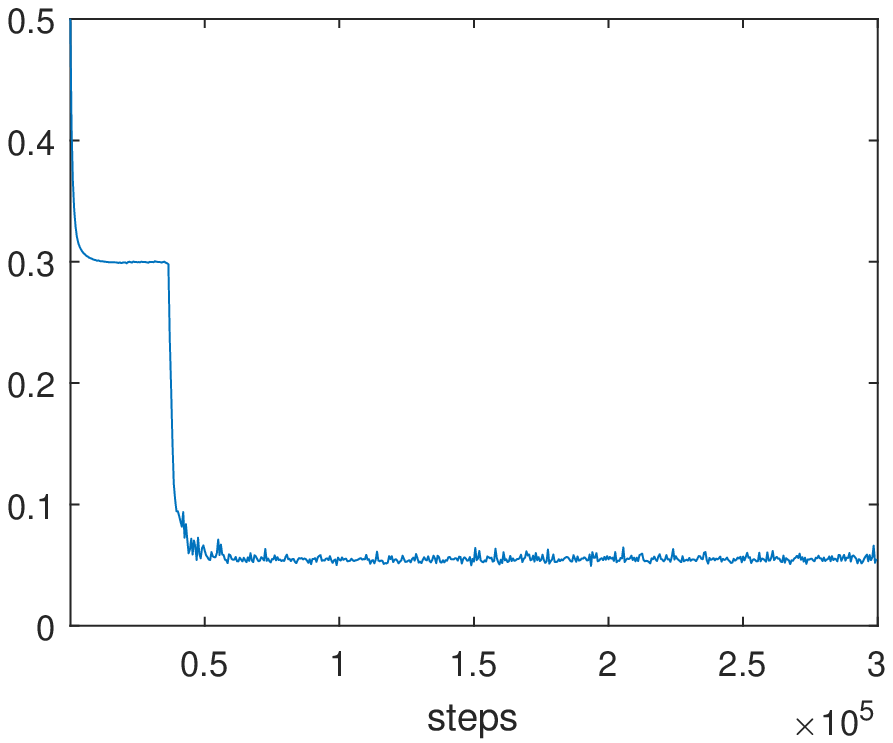}
	\caption{}
	\label{fig:eg1-f}
    \end{subfigure} \\ 
	\caption{High contrast problem, $\alpha_0=1000$, $\alpha_1=1$ case: 
		(a) profile of $g$;
		(b) profile of $u'$;
		(c) decay of the Lagrangian during the training process; 
		(d) profile of the DNN solution $u$ at the final step;
		(e) profile of the reference solution; 
		(f) decay of the $L_2$ relative error during the training process. }
	\label{fig:eg1}
\end{figure}

In Fig.\ref{fig:eg1}, we show the corresponding numerical results. In Fig.\ref{fig:eg1-a} and Fig.\ref{fig:eg1-b}, we plot the profiles of a shallow network approximation of the boundary condition $g(x)$ and the deep network approximation of solution $u'(x)$ to the auxiliary PDE \eqref{meth:nonzero-diri}, respectively. In Fig.\ref{fig:eg1-d} and Fig.\ref{fig:eg1-e}, we show the comparison between the DNN solution and the reference solution. One can see that the DNN method provides an accurate result for this interface problem. 

In Fig.\ref{fig:eg1-c} and Fig.\ref{fig:eg1-f}, we plot the decay of the Lagrangian and the $L_2$ relative error between the DNN solution and reference solution during the training process. Interestingly we observe that optimization process gets stuck at a local minimum at the beginning, i.e., the first four thousand steps, where the Lagrangian functional does not have decay and the error between the DNN solution and reference solution keeps as a constant. Beyond that the optimization process jumps out the local minimum, which make
the Lagrangian functional and the error continue to decay. Finally the error oscillates around 5\%. 
 
In our second experiment, we choose $\alpha_0=1$ and $\alpha_1=10^3$ in \eqref{NumEllipCoef}. The profile of the new coefficient looks like an upside down of the profile shown in Fig.\ref{fig:eg1alpha3d}. We do not show it here. Again, we use the immersed-interface FEM with fine mesh  $h=\frac{1}{128}$ to compute the reference solution and the DNN method to compute the numerical solution. The setting of the DNN method is the same as the first experiment. 

\begin{figure}
	\centering
	\begin{subfigure}[b]{0.31\textwidth}
		\includegraphics[width=\textwidth]{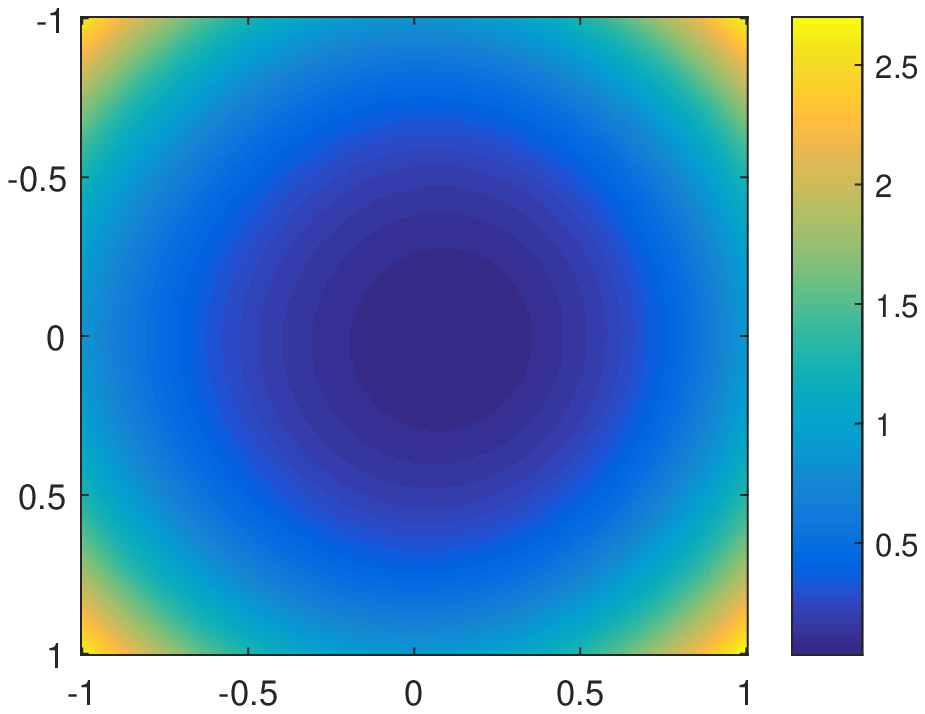}
		\caption{}
		\label{fig:eg3-a}
	\end{subfigure}
	~ %add desired spacing between images, e. g. ~, \quad, \qquad, \hfill etc. 
	\begin{subfigure}[b]{0.31\textwidth}
		\includegraphics[width=\textwidth]{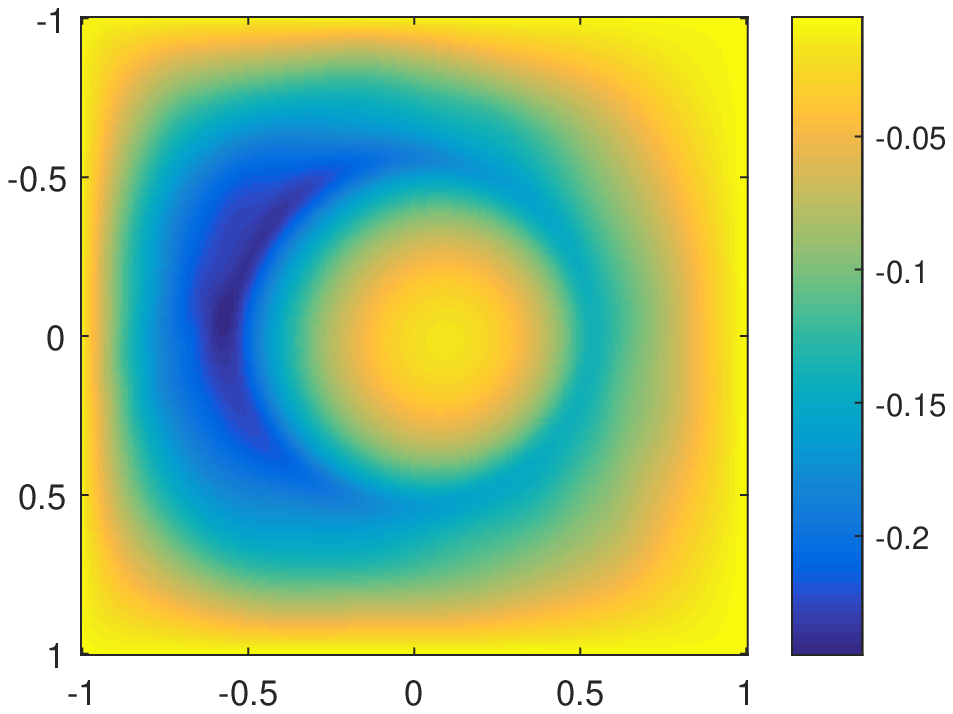}
		\caption{}
		\label{fig:eg3-b}
	\end{subfigure}
	~ 
	\begin{subfigure}[b]{0.31\textwidth}
		\includegraphics[width=\textwidth]{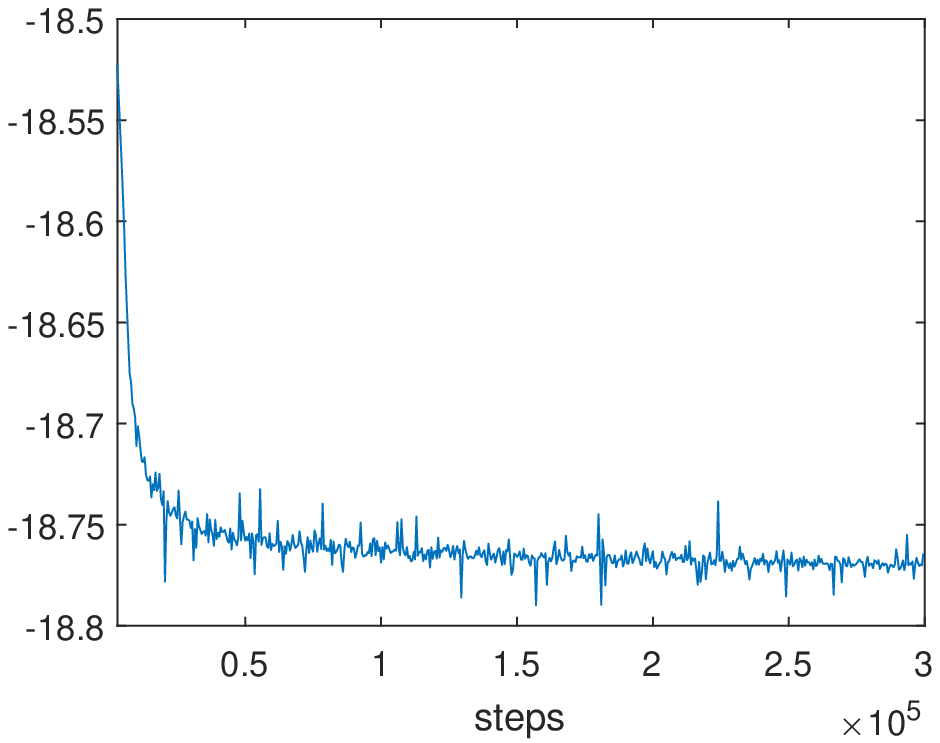}
		\caption{}
		\label{fig:eg3-c}
	\end{subfigure} \\ 
	\centering
	\begin{subfigure}[b]{0.31\textwidth}
		\includegraphics[width=\textwidth]{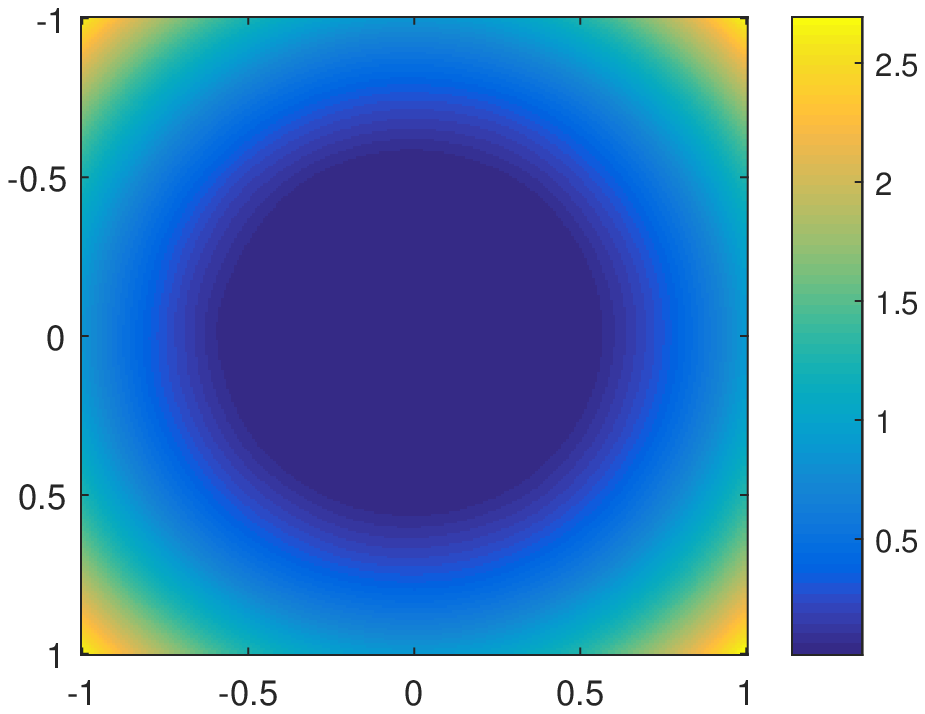}
		\caption{}
		\label{fig:eg3-d}
	\end{subfigure}
	~ %add desired spacing between images, e. g. ~, \quad, \qquad, \hfill etc. 
	\begin{subfigure}[b]{0.31\textwidth}
		\includegraphics[width=\textwidth]{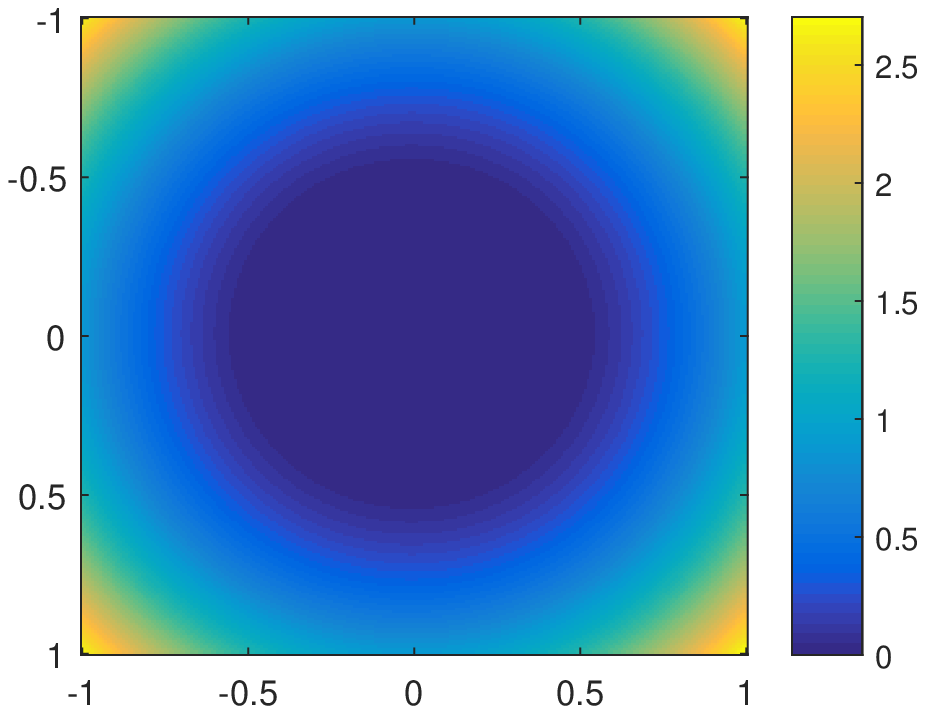}
		\caption{}
		\label{fig:eg3-e}
	\end{subfigure}
	~ 
	\begin{subfigure}[b]{0.31\textwidth}
		\includegraphics[width=\textwidth]{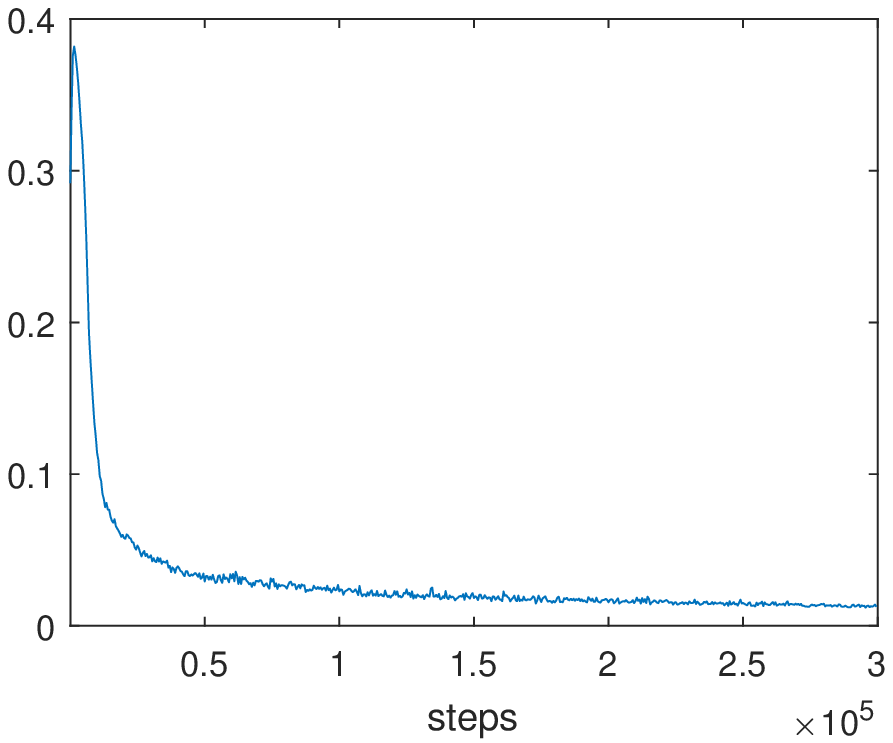}
		\caption{}
		\label{fig:eg3-f}
	\end{subfigure} \\ 
	\caption{High contrast problem, $\alpha_0=1$, $\alpha_1=1000$ case: 
		(a) profile of $g$;
		(b) profile of $u'$;
		(c) decay of the Lagrangian during the training process; 
		(d) profile of the DNN solution $u$ at the final step;
		(e) profile of the reference solution; 
		(f) decay of the $L_2$ relative error during the training process. }
	\label{fig:eg3}
\end{figure}

In Fig.\ref{fig:eg3}, we show the corresponding numerical results. In Fig.\ref{fig:eg3-a} and Fig.\ref{fig:eg3-b}, we plot the profiles of a shallow network approximation of the boundary condition $g(x)$ and the deep network approximation of solution $u'(x)$ to the auxiliary PDE \eqref{meth:nonzero-diri}, respectively. In Fig.\ref{fig:eg3-d} and Fig.\ref{fig:eg3-e}, we show the comparison between the DNN solution and the reference solution. The DNN method also provides an accurate result for this interface problem. 

In Fig.\ref{fig:eg3-c} and Fig.\ref{fig:eg3-f}, we plot the decay of the Lagrangian and the $L_2$ relative error between the DNN solution and reference solution during the training process.  We find that the decay pattern of the second experiment is different from the first one. The Lagrangian functional has instant fluctuations during the optimization process. However, it does not get stuck at a local minimum.  The error function is a monotonic decreasing function. Finally the error is reduced to about 2\%.

\begin{figure}
	\centering
	\begin{subfigure}[b]{0.45\textwidth}
		\includegraphics[width=\textwidth]{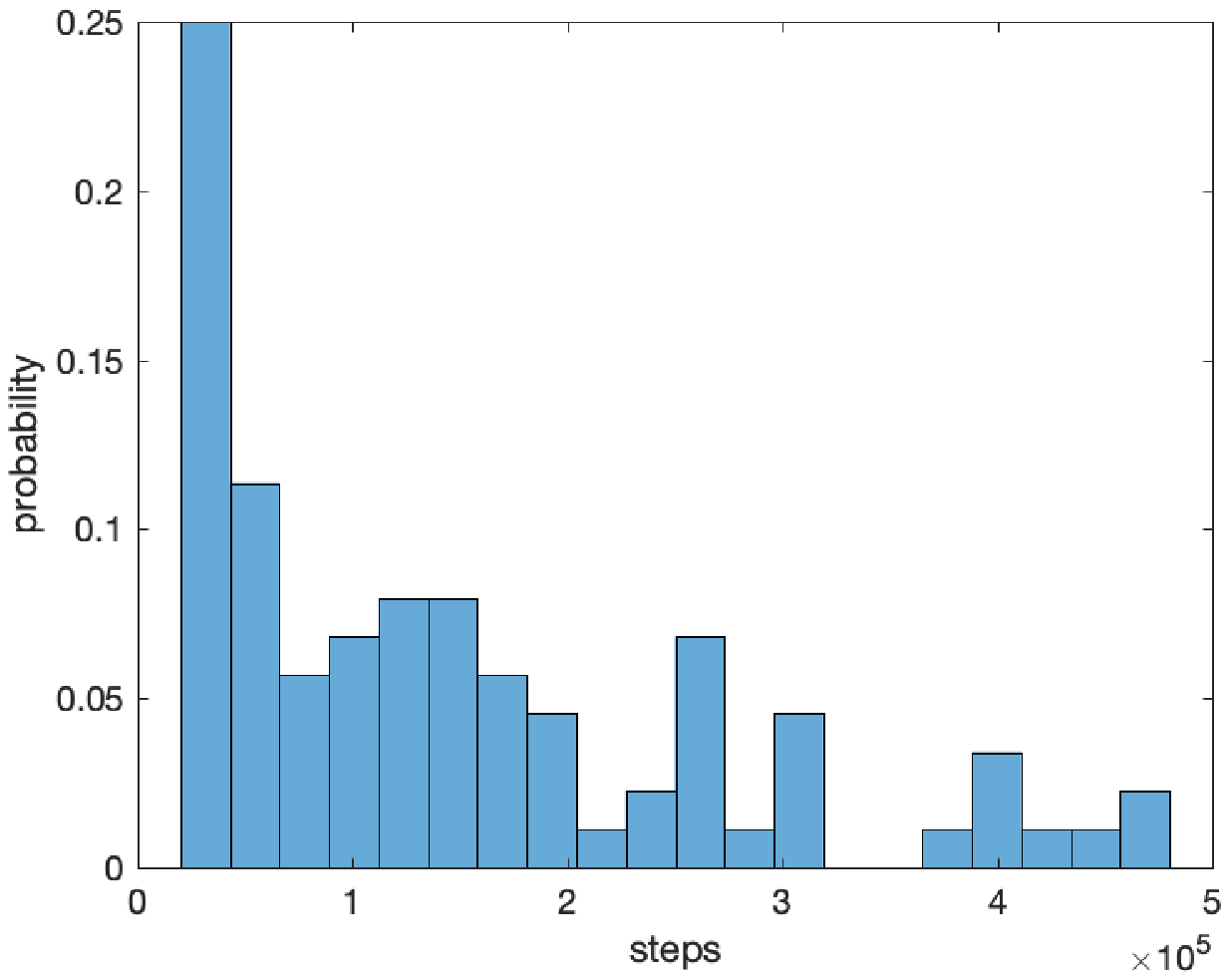}
		\caption{$\alpha_0=1000$, $\alpha_1=1$.}
		\label{fig:mctest-a}
	\end{subfigure}
	~ %add desired spacing between images, e. g. ~, \quad, \qquad, \hfill etc. 
	\begin{subfigure}[b]{0.45\textwidth}
		\includegraphics[width=\textwidth]{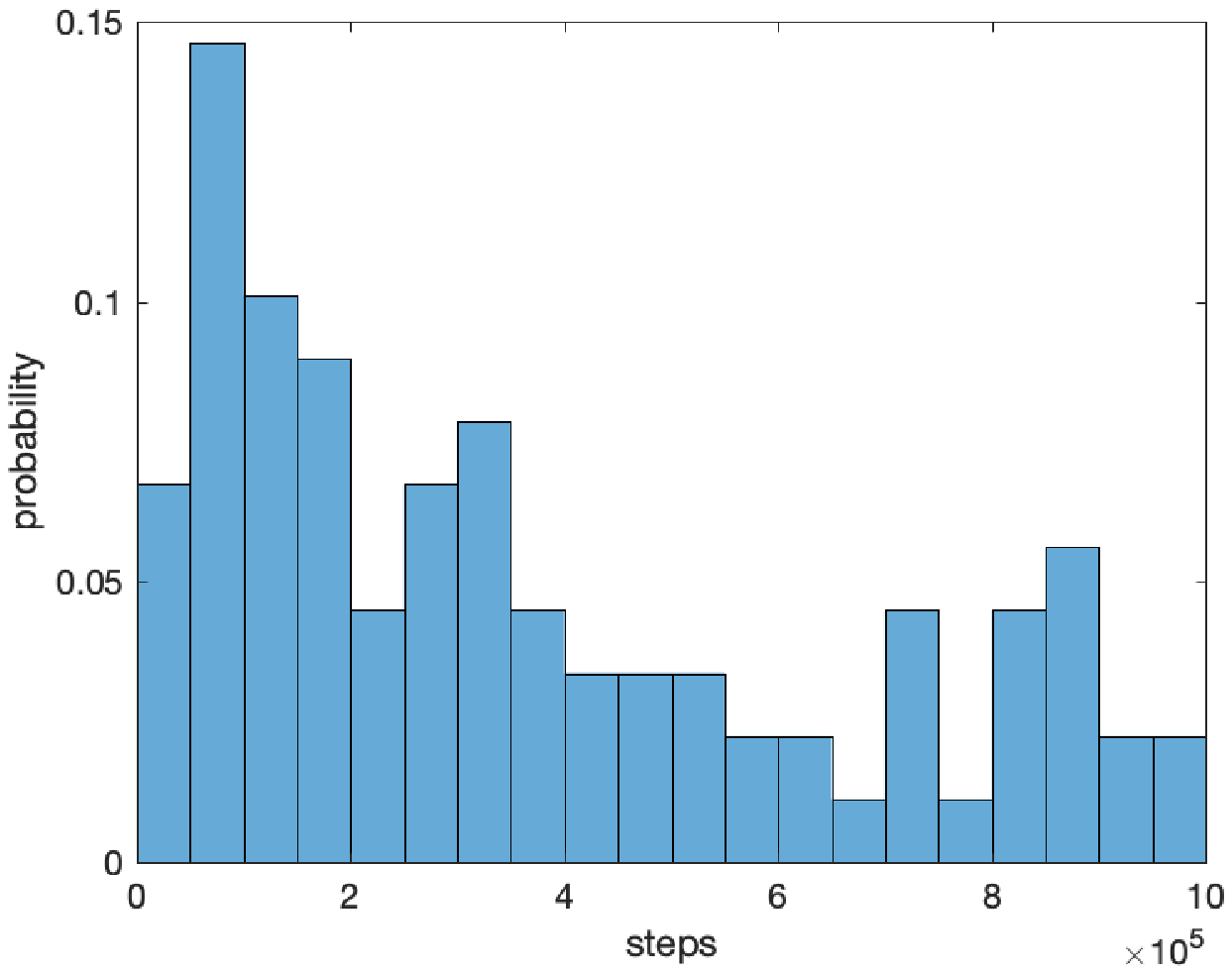}
		\caption{$\alpha_0=10000$, $\alpha_1=1$.}
		\label{fig:mctest-b}
	\end{subfigure}
	\caption{Histogram of the number of steps to obtain a convergence result.}
	\label{fig:eg1_mctest}
\end{figure}

The DNN method is a probabilistic method since the initial value of parameters in the network, 
i.e. $\theta\in\Theta$ and the Adams SGD optimizer are random. We are interested in investigating 
the convergence speed when $\alpha_1=1$ and $\alpha_0\gg 1$, which is a `harder' case of the high-contrast problem since the optimization process of the DNN method gets stuck at a local minimum. In Fig.\ref{fig:eg1_mctest}, we show results of the convergence speed study when $\alpha_0=1000$ and  $\alpha_0=10000$, respectively. Specifically, we plot the histogram of the number of steps to converge. The total number of iteration is $5\times10^5$ when $\alpha_0=1000$ and $10^6$ when $\alpha_0=10000$. We find that a higher contrast in the coefficient will lead to a slower convergence in the DNN method.  We also find that about $7\%$ of trials failed to converge 
within the designed steps.
 
\subsection{2D Linear elasticity interface problem}
\noindent
We consider a linear elasticity PDE with a discontinuous stress tensor as follows, 
\begin{equation}\label{eqn:num_compact_elastic}
-\nabla\cdot\big(\mathbb{C}\nabla \textbf{u}+\chi_{\Omega}S_0\big)=0, \quad  x\in D, 
\end{equation}
where $x=(x_1,x_2)$, the domain $D=[-8,8]\times[-8,8]$, $\textbf{u}=(u_1,u_2)^{T}$, the elasticity tensor $\mathbb{C}$ is defined by \eqref{ElasticityTensor1} or \eqref{ElasticityTensor2} with $\lambda=1$ and 
$\mu=1$. %This setting is equivalent to poisson ratio $\nu = 1/4$ and Young's modulus $E = 1$.

\begin{figure}
	\centering
	\includegraphics[width=0.6\linewidth]{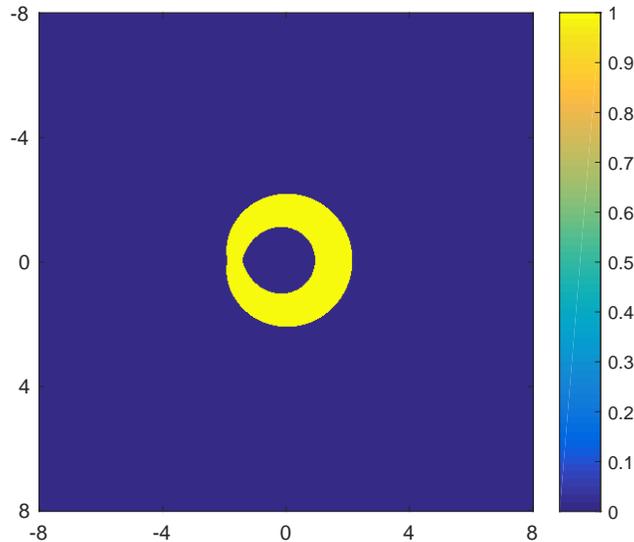}
	\caption{Value of $\chi_\Omega$ on $D$, where the yellow region is $\Omega$.}
	\label{fig:eg2chi}
\end{figure}

In the cell model \cite{zhang2018minimal}, keratocytes typically have a roughly circular shape with an annular lamellipodium surrounding the nucleus, when they are in stationary state. Contact and force transmission with the substrate occurs only at the lamellipodium and not the nucleus and organelles. Accordingly, we choose the initial lamellipodium region $\Omega$ to be an annulus in the center of the square domain $D$, with the nucleus excluded; see Fig.\ref{fig:eg2chi}. 

We set $u_1=u_2=0$ on the boundary of $D$, which gives a null displacement or traction-free boundary condition. On the boundary of the cell $\Omega$, we impose the jump conditions \eqref{eqn:jump_conditions}.

We use the immersed-interface FEM with a fine mesh $h=1/32$ to compute the reference solution and the DNN method to compute the numerical solution. The network maps $x\in\mathbb{R}^2$ to $\textbf{u}\in\mathbb{R}^2$ which used 4 intermediate layers. The width of each layer is 20 and layout is same with Fig.\ref{twoDNNs_u}. In the running of the SGD method, we choose the batch number to be $2048$ %(including $256$ points on the boundary) 
and generate a new batch every $10$ steps of updating. And the learning rate $\eta$ is $5\times 10^{-4}$.

\begin{figure}
	\centering
	\begin{subfigure}[b]{0.31\textwidth}
		\includegraphics[width=\textwidth]{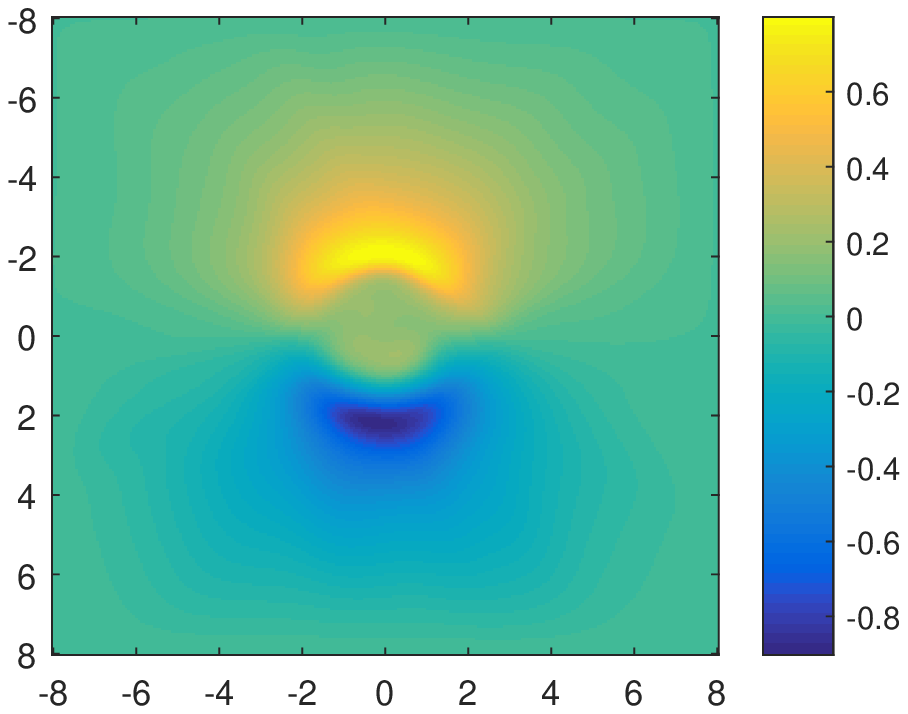}
		\caption{}
		\label{fig:eg2-a}
	\end{subfigure}
	~ %add desired spacing between images, e. g. ~, \quad, \qquad, \hfill etc. 
	\begin{subfigure}[b]{0.31\textwidth}
		\includegraphics[width=\textwidth]{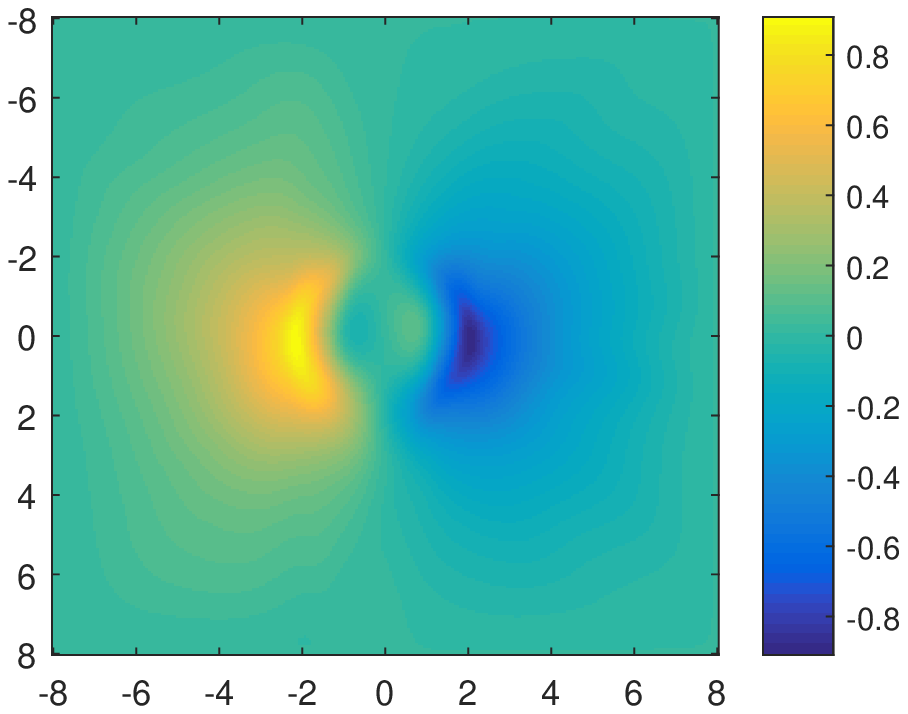}
		\caption{}
		\label{fig:eg2-b}
	\end{subfigure}
	~ 
	\begin{subfigure}[b]{0.31\textwidth}
		\includegraphics[width=\textwidth]{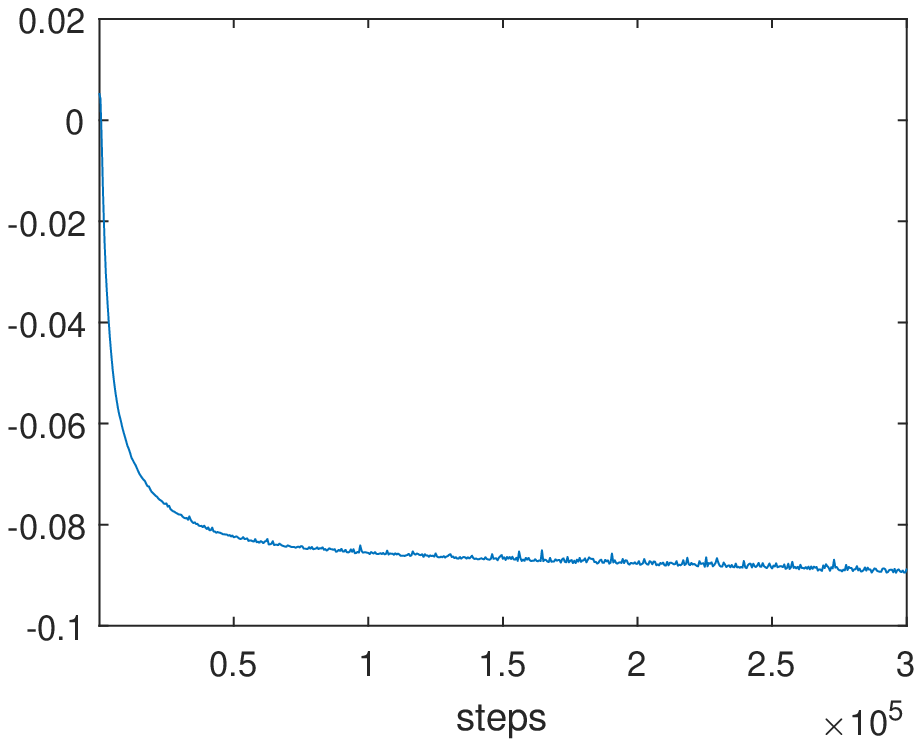}
		\caption{}
		\label{fig:eg2-c}
	\end{subfigure} \\ 
	\centering
	\begin{subfigure}[b]{0.31\textwidth}
		\includegraphics[width=\textwidth]{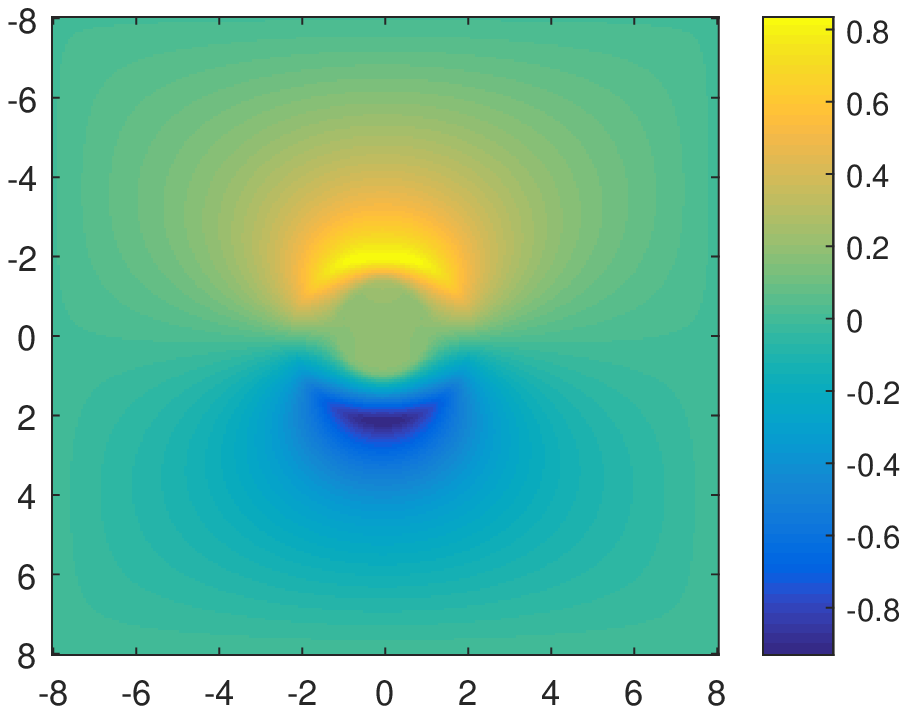}
		\caption{}
		\label{fig:eg2-d}
	\end{subfigure}
	~ %add desired spacing between images, e. g. ~, \quad, \qquad, \hfill etc. 
	\begin{subfigure}[b]{0.31\textwidth}
		\includegraphics[width=\textwidth]{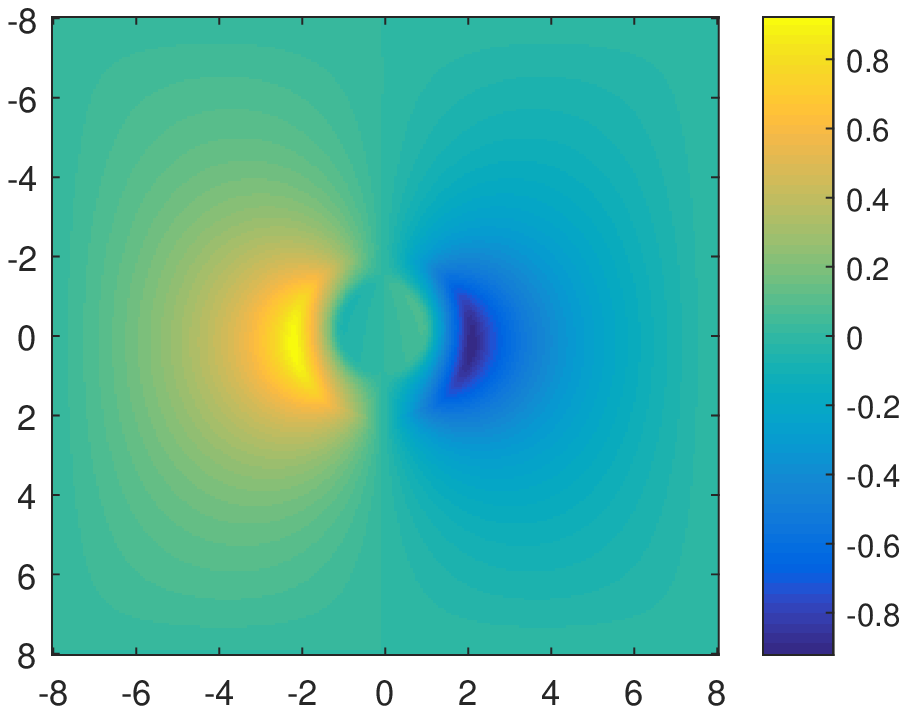}
		\caption{}
		\label{fig:eg2-e}
	\end{subfigure}
	~ 
	\begin{subfigure}[b]{0.31\textwidth}
		\includegraphics[width=\textwidth]{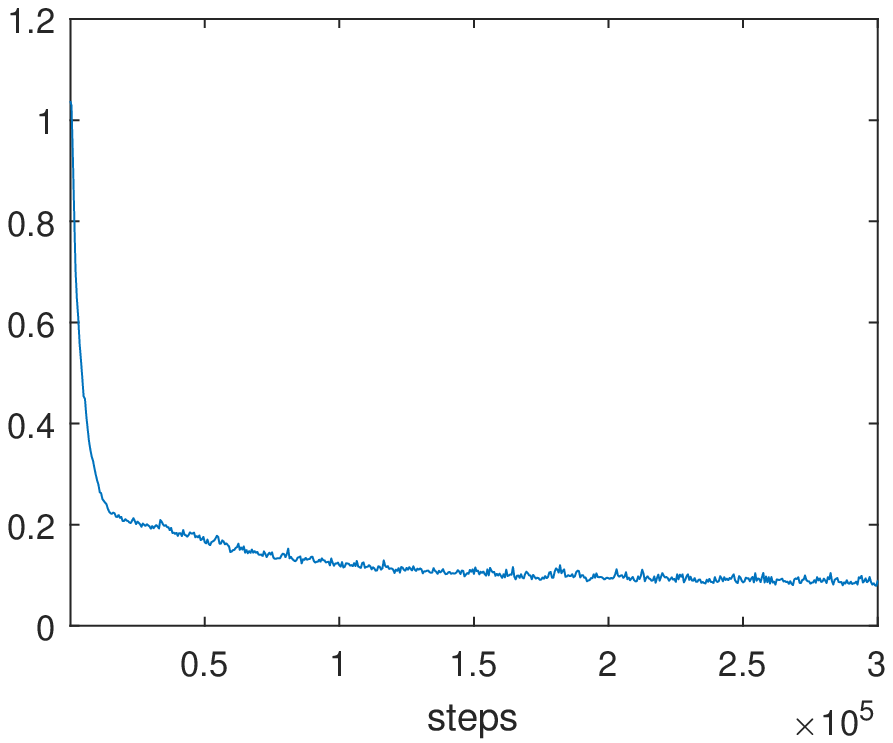}
		\caption{}
		\label{fig:eg2-f}
	\end{subfigure} \\ 
	\caption{2D Linear elasticity interface problem: 
		(a) profile of DNN solution $u_1$;
		(b) profile of DNN solution $u_2$;
		(c) decay of the Lagrangian during the training process; 
		(d) profile of reference solution $u_1$;
		(e) profile of reference solution $u_2$; 
		(f) decay of the $L_2$ relative error during the training process. }
	\label{fig:eg2}
\end{figure}

In Fig.\ref{fig:eg2}, we show the corresponding numerical results. In Fig.\ref{fig:eg2-a} and Fig.\ref{fig:eg2-b}, we plot the profiles of DNN solutions $u_1$ and $u_2$, which are the displacements in $x_1$ and $x_2$ coordinates, respectively. The corresponding reference solutions are shown in Fig.\ref{fig:eg2-d} and Fig.\ref{fig:eg2-e}. We find that the DNN solutions agree well with the reference solutions. In Fig.\ref{fig:eg2-c} and Fig.\ref{fig:eg2-f}, we plot the decay of the Lagrangian and the $L_2$ relative error between the DNN solution and reference solution during the training process. We find that the decay pattern of the third experiment is same as the second one.   Finally the error is reduced to about 4\%. Our numerical results imply that the DNN method is efficient in solving the 2D Linear elasticity interface problem \eqref{eqn:num_compact_elastic}. Most importantly, its implementation is very simple.    

\section{Conclusions}
\noindent
In this paper, we studied the deep-learning based method to solve interface problems. By formulating the PDEs into variational problems, we convert the interface problems into optimization problems. Since the DNN can be used to approximate the linear space spanned by FEM nodal basis functions. Thus, we parameterize the PDE solutions using the DNN and solve the interface problems by searching the minimizer of the associated optimization problems. Although the parameter space of the DNN is huge, the SGD method can be applied to solve 
the optimization problems efficiently. 
%which happened in many other classic deep-learning algorithms for wide range of actual problems. 
In this framework, once we have samplers of grids on the domain and the boundary, we do not need any special treatment to deal with the interface inside the domain. Therefore, the proposed method is easy to implement and mesh-free.  Finally, we present numerical experiments to demonstrate the performance of the proposed method. Specifically, we use the DNN method to solve elliptic PDEs with discontinuous and high-contrast coefficients and linear elasticity with discontinuous stress tensors. 
We find the the DNN method gives accurate results for both experiments. There are several issues remain open. For instance, we do not get the convergence rate for the DNN method and we have little understanding about the parameter space of the DNN. In addition, the issue of local minima and saddle points in the optimization problem is highly nontrivial. We are interested in studying these issues in our future research.

\section*{Acknowledgements}
\noindent
The research of Z. Wang is partially supported by the Hong Kong PhD Fellowship Scheme. The research of Z. Zhang is supported by Hong Kong RGC grants (Projects 27300616, 17300817, and 17300318), National Natural Science Foundation of China (Project 11601457), Seed Funding Programme for Basic Research (HKU), an RAE Improvement Fund from the Faculty of Science (HKU), and the Hung Hing Ying Physical Sciences Research Fund (HKU). The computations were performed using the HKU ITS research computing facilities that are supported in part by the Hong Kong UGC Special Equipment Grant (SEG HKU09). We would like to thank Professor Thomas Hou for stimulating discussions.

\appendix

\section*{References}
\bibliographystyle{plain}
\bibliography{ZWpaper_DeepLearningPDEDiscCoef}

\end{document}